# METRIPLECTIC FORMALISM: FRICTION AND MUCH MORE

Massimo Materassi
Istituto dei Sistemi Complessi
Consiglio Nazionale delle Ricerche
Italy
massimo.materassi@isc.cnr.it

Philip J. Morrison
Physics Department & IFS
The University of Texas at Austin
United States of America
morrison@physics.utexas.edu

**Abstract**

The metriplectic formalism [Morrison, 1984] couples Poisson brackets of the Hamiltonian description with metric brackets for describing systems with both Hamiltonian and dissipative components. The construction builds in asymptotic convergence to a preselected equilibrium state. Phenomena such as friction, electric resistivity, thermal conductivity and collisions in kinetic theories are well represented in this framework. In this paper we present an application of the metriplectic formalism of interest for the theory of control: a suitable torque is applied to a free rigid body, which is expressed through a metriplectic extension of its "natural" Poisson algebra. On practical grounds, the effect is to drive the body to align its angular velocity to rotation about a stable principal axis of inertia, while conserving its kinetic energy in the process. On theoretical grounds, this example shows how the non-Hamiltonian part of a metriplectic system may include convergence to a limit cycle, the first example of a non-zero dimensional attractor in this formalism. The method suggests a way to extend metriplectic dynamics to systems with general attractors, e.g. chaotic ones, with the hope of representing bio-physical, geophysical and ecological models.

## 1 Introduction

In this work an extension of the Hamiltonian formalism is considered, one that is able to include "dissipation" and the asymptotic convergence to solutions. This is realized by adding to the Poisson bracket a semi-metric bracket. The resulting Leibniz algebra governing the evolution is a *metriplectic bracket algebra*.

In particular, the class of *complete metriplectic systems* (CMS) is considered, in which the total energy of the system, namely the Hamiltonian $H$, is conserved, while another functional, referred to as the entropy $S$, grows until its maximum, which corresponds to an asymptotic equilibrium state, is reached.

Complete metriplectic systems have been written for many cases in which a Hamiltonian physical system is coupled to "microscopic degrees of freedom" causing *friction* or *thermal conduction* [Materassi and Tassi, 2012]; in the case of plasma kinetic theory, "higher order" terms collected in collision integrals are recast in a semi-metric guise, so as to render the collisional Poisson-Vlasov system a complete metriplectic system [Morrison, 1984].

Here we focus on a particular case, already introduced in [Morrison, 1986], in which the Hamiltonian system representing a free rigid body is perturbed with an external torque $\vec{\tau}_{\text{servo}}$ suitably *designed to modify the angular momentum $\vec{L}$ without changing the energy of the system*. In particular, the action of this torque makes the angular velocity $\vec{\omega}$ converge to a free rotation around one principal axis. This can be mathematically accomplished by adding to the Poisson algebra a semi-metric term based on the invariance of $L^2$ under the action of the Poisson bracket of the Hamiltonian part of the system. As the torque applied must be a suitable function of the angular momentum $\vec{\tau}_{\text{servo}} = \vec{\tau}_{\text{servo}}(\vec{L})$, the technological solution imagined is a servo-motor, indicated as *metriplectic servo-motor* (MSM) in what follows. The intriguing aspect is that, due to the energy conservation, the MSM would re-direct $\vec{\omega}$ without any power consumption, as $\vec{\tau}_{\text{servo}} \cdot \vec{\omega} = 0$.

The paper is organized as follows.

In § 2 a short review of the metriplectic algebra is presented, and some remarkable examples of CMS are quoted. In § 3 the application of the MSM to the free rigid body is described, first using the angular velocity space as the phase space of the system, then describing its metriplectic dynamics through the canonical variables $(\chi, \vec{p})$, with $\chi$ being Euler angles and $\vec{p}$ their canonically conjugate momenta. It is shown that the rotation around a principal axis of inertia corresponds to an asymptotic equilibrium point in the $\vec{\omega}$ space, while it is a linearly stable orbit in the canonical variables $(\chi, \vec{p})$.

Conclusions and further development of the present study, both in physical and technological senses, are drawn in § 4.

**2 Metriplectic systems: a review**

Complete metriplectic systems are dynamical systems governed by an extension of the usual Poisson brackets of the Hamiltonian systems. Typically, one starts with a Hamiltonian system of Hamiltonian $H(z)$ and Poisson bracket $\{.,.\}$, $z$ being the dynamical variables that are coordinates of the phase space (for canonical systems $z = (q, p)$, the conjugate pair, which is not the most general Hamiltonian case [Morrison, 1998]). Hamiltonian dynamics is governed by $\{.,.\}$ and $H(z)$, such that any observable $A$ has dynamics given by:

$$\dot{A}(z) = \{A(z), H(z)\} \tag{1}$$

In general, this dynamics conserves $H$, due to the anti-symmetric property of $\{.,.\}$, plus a certain number of other quantities may be conserved. It may be the case that some quantity $S$ exists, such that it has null Poisson bracket with *any* other function of $z$:

$$\{S(z), A(z)\} = 0 \ \forall \ A.$$

Such a quantity, referred to as a *Casimir of the Poisson bracket*, is necessarily conserved by the dynamics (1) independent of the particular Hamiltonian. If some semi-metric Leibniz bracket $(.,.)$ is defined

$$(A, B) = (B, A), (A, A) \leq 0 \ \forall A, B, \tag{2}$$

such that $H$ is zero with any other observable

$$(A, H) = 0 \ \forall A \tag{3}$$

then the Leibniz bracket dynamics

$$\dot{A}(z) = \{A(z), H(z)\} + \zeta(A(z), S(z)) \tag{4}$$

will *conserve the Hamiltonian* because of (3) and *increase the Casimir S* because of (2). If the *metriplectic bracket*

$$\langle\langle A, B \rangle\rangle = \{A, B\} + (A, B)$$

is defined, then, provided the *free energy functional* $F = H + \zeta S$ is used, the *complete metriplectic dynamics* may be constructed:

$$\dot{A}(z) = \langle\langle A(z), F(z) \rangle\rangle; \tag{5}$$

this dynamics *conserves the Hamiltonian and increases the Casimir*

$$\dot{H}(z) = 0, \dot{S}(z) \geq 0 \tag{6}$$

and the second condition in (6) is realized as the constant $\zeta$ in $F = H + \zeta S$ is negative. The *asymptotic equilibria of (5) are the extrema of F*, the Casimir $S$ playing the role of *a Lyapunov functional*. Moreover, thanks to $\dot{S}(z) \geq 0$, such a dynamics is *irreversible* in the thermodynamic sense.

Essentially on this basis, the perturbation of a Hamiltonian system with *dissipative interactions*, draining irreversibly its energy to some microscopic degrees of freedom, may give rise to a CMS of the form (5), where $H$ represents the *total energy*, including the thermal energy of the microscopic degrees of freedom, while $S$ is *the entropy* of the closed system made by the formerly Hamiltonian one plus the thermal bath giving rise to friction. The functional $F$ is definitely interpreted as free energy, in this case.

The aforementioned characteristics render CMS ideal for representing systems that are Hamiltonian in the limit of no dissipation and no thermal conduction, that relax to asymptotic equilibria due to the interaction of "macroscopic", deterministic degrees of freedom with "microscopic" degrees of freedom (represented via thermodynamics). This is the case of mechanical systems with friction, dissipative hydrodynamics and visco-resistive magnetohydrodynamics: in all those examples, the dynamical variables $z$ are a collection of "mechanical" variables (e.g., the position and momentum of a particle, bulk velocity of a fluid/plasma, magnetic field) plus variables representing the microscopic degrees of freedom draining irreversibly energy via dissipative processes. Another example of CMS is the collisional Vlasov-Poisson equation, in which the Hamiltonian system of a kinetic theory of a plasma without collisions is metriplectically extended with the inclusion of the collision integral of the Landau-Lenard-Bălescu form: in this case, there are no "macroscopic" and "microscopic" degrees of freedom, rather single or multiple particle processes.

In § 3 of the present paper, a particular application of the CMS is presented, where the Casimir functional $S$ does not depend on degrees of freedom other than those of the Hamiltonian system. This tricky interesting example is that of a *free rigid body*.

**3 The metriplectic servo-motor for the rigid body**

Consider a rigid body with inertia tensor $\sigma$ with the three eigenvalues $\{I_1, I_2, I_3\}$. The rigid body may be described via a phase space spanned by canonical variables, in particular, the set $(\chi, \vec{p})$ of the three *Euler angles* $\chi$ and their canonical momenta $\vec{p}$ (in our notation, only the triplet $\vec{p}$ is indicated as a vector, because it is a spin-1 representation of the rotation group $SO(3)$, while the collection $\chi$ of the three Euler angles do not transform as a vector under $SO(3)$). In the absence of external torques, the Hamiltonian of the system reduces to its kinetic energy only

$$H(\chi, \vec{p}) = \frac{1}{2} \vec{L}^{\mathrm{T}}(\chi, \vec{p}) \cdot \sigma^{-1} \cdot \vec{L}(\chi, \vec{p}), \tag{7}$$

with $\vec{L}$ being the angular momentum of the rigid body. The three components of $\vec{L}$ form *a closed non-canonical Poisson algebra*, given by

$$\{L_i, L_j\} = -\varepsilon_{ij}{}^k L_k \tag{8}$$

(summation is intended over repeated indices), i.e. the symplectic realization of the Lie algebra of rotations *so*(3). It is hence possible *to reduce* the free rigid body Hamiltonian dynamics [Morrison, 1998], originally given by (7) and $\{\chi^i, p_j\} = \delta^i_j$, to a dynamics all expressed in terms of the components of $\vec{L}$, with Hamiltonian

$$H(\vec{L}) = \frac{1}{2} \vec{L}^{\mathrm{T}} \cdot \sigma^{-1} \cdot \vec{L} \tag{9}$$

and Poisson algebra (8). The Hamiltonian motion of the free rigid body is determined as

$$\dot{A}(\vec{L}) = \{A(\vec{L}), H(\vec{L})\}. \tag{10}$$

Given (8), any function $C(L^2)$ of the square modulus of $\vec{L}$ is a Casimir, since

$$\{C(L^2), A(\vec{L})\} = 0 \, \forall \, A.$$

The construction of a CMS out of the Hamiltonian system (9) and (8) is as follows: a certain symmetric semi-definite tensor $\Gamma(\vec{L})$ is constructed such that $\Gamma(\vec{L}) \cdot \frac{\partial H}{\partial \vec{L}} = 0$; then the symmetric bracket is defined as

$$(A, B) = \Gamma^{ij} \frac{\partial A}{\partial L^i} \frac{\partial B}{\partial L^j},$$

so that the dynamics

$$\dot{A}(\vec{L}) = \{A(\vec{L}), H(\vec{L})\} + \zeta(A(\vec{L}), C(L^2)) \tag{11}$$

represents a CMS. The simplest possible choice for the tensor $\Gamma(\vec{L})$ is

$$\Gamma(\vec{L}) = \frac{k}{\zeta}(\omega^2 \mathbf{1} - \vec{\omega} \otimes \vec{\omega}) \quad / \quad \vec{\omega} = \sigma^{-1} \cdot \vec{L}$$

($\vec{\omega}$ being the *angular velocity* of the rigid body). This dynamics conserves $H$ and increases $C$ monotonically, until the total free energy

$$F(\vec{L}) = H(\vec{L}) + \zeta C(L^2) \tag{12}$$

reaches its extremum, $\frac{\partial F}{\partial \vec{L}} = 0$: this condition is realized for $\vec{\omega} = -2\zeta C'(L^2)\vec{L}$, being $C' = \frac{dC}{dL^2}$, i.e. when the angular velocity of the rigid body and its angular momentum $\vec{L}$ *are aligned*. In other words, the dynamics (11) drives the rigid body *to relax* to the condition of *rotation along one (stable) principal axis of inertia*:

$$\vec{\omega}_{eq} = I_{eq} \vec{L}_{eq} \quad / \quad I_{eq} = -2\zeta C'(L^2_{eq}) \tag{13}$$

(clearly, as $I_{eq}$ is a positive quantity, the constant $\zeta$ is to be taken negative, as one takes the function $C(L^2)$ increasing with $L^2$).

The dynamics (11) changes the vector $\vec{L}$ (without changing the energy $H(\vec{L})$), so it must be equivalent to the application of an external torque. In particular, one may see that the torque encoded in (11) is:

$$\vec{\tau}_{servo}(\vec{L}) = 2kC'(L^2)[\omega^2 \vec{L} - (\vec{\omega} \cdot \vec{L})\vec{\omega}]. \tag{14}$$

This torque must be applied from outside, but still has to depend on the instantaneous motion of the rigid body through $\vec{L}$ and $\vec{\omega}$: in order to *realize technologically* this system one has to use *a servo-mechanism*, constantly reading what the rigid body is precisely doing. Such a mechanism will be referred to as *metriplectic servo-motor* (MSM). Remarkably, as one may see

$$\vec{\tau}_{servo}(\vec{L}) \cdot \vec{\omega} = 0,$$

the *mechanical* power of the servo-motor vanishes, so that such a MSM could drive to alignment (13) a rigid body *of any size* with no power at all, as far as the mechanical labor is concerned. Of course, as $\vec{\tau}_{servo}$ depends on the condition of the rigid body, it must take some energy to measure, save and react to this.

In the phase space of the vectors $\vec{L}$ or $\vec{\omega}$, the *reduced phase space*, the equilibrium to which the CMS (11) tends is a *point-like asymptotic equilibrium*, namely a condition of *entropic death* after which the system doesn't evolve any more. This is identical to the examples of CMS quoted in § 2.

The equations of motion of $\vec{L}$ form the MSM

$$\dot{\vec{L}} = \vec{L} \times (\sigma^{-1} \cdot \vec{L}) +$$
$$+ 2kC'(L^2)[(\vec{L}^T \cdot \sigma^{-2} \cdot \vec{L})\vec{L} - (\vec{L}^T \cdot \sigma^{-1} \cdot \vec{L})(\sigma^{-1} \cdot \vec{L})].$$

It is rather easy to see that when $\vec{L}$ lies along an eigen-direction of $\sigma^{-1}$, i.e. of $\sigma$, the foregoing dynamics vanishes. In particular, it is also possible to show that, if the eigenvalues of $\sigma$ are ordered as $I_1 > I_2 > I_3$, only the states $\vec{L}_{(1)} = (L,0,0)^T$ and $\vec{L}_{(3)} = (0,0,L)^T$ are stable equilibria, while $\vec{L}_{(2)} = (0,L,0)^T$ is unstable, for any real value *L*.

An important characteristic of the metriplectic formalism is that the metric part $\zeta(A(z), S(z))$ of the dynamics, responsible for "relaxation", may indeed be designed so to make the system relax to one stable equilibrium instead of another. Let us give a practical example here of this mechanism by working with the metric part $2kC'(L^2)[(\vec{L}^T \cdot \sigma^{-2} \cdot \vec{L})\vec{L} - (\vec{L}^T \cdot \sigma^{-1} \cdot \vec{L})(\sigma^{-1} \cdot \vec{L})]$ of the MSM for the rigid body. In particular: the Casimir $C(L^2)$ can either be designed so to make the system relax to $\vec{L}_{(1)} = (L,0,0)^T$, or to $\vec{L}_{(3)} = (0,0,L)^T$.

In the following Figures 1-6 we have reported the time series $\vec{L}(t)$ and the phase portrait, in the *L*-space, of the MSM with $C(L^2)$ designed so that $2kC'(L^2) = 0.1$, for different initial conditions, imitating the proximity of presumably equilibrium points, with a rigid body of momenta of inertia equal to $I_1 = 10$, $I_2 = 5$ and $I_3 = 1$ in arbitrary units. In the Figures 7-12, instead, the condition on $C(L^2)$ is changed so that $2kC'(L^2) = -0.1$, while all the remaining parameters, and initial conditions, are kept the same. We see numerically that, while for $2kC'(L^2) = 0.1$ the system relaxes to $\vec{L}_{(1)} = (L,0,0)^T$, setting $2kC'(L^2) = -0.1$ the state $\vec{L}_{(3)} = (0,0,L)^T$ turns out to be the global attractor.

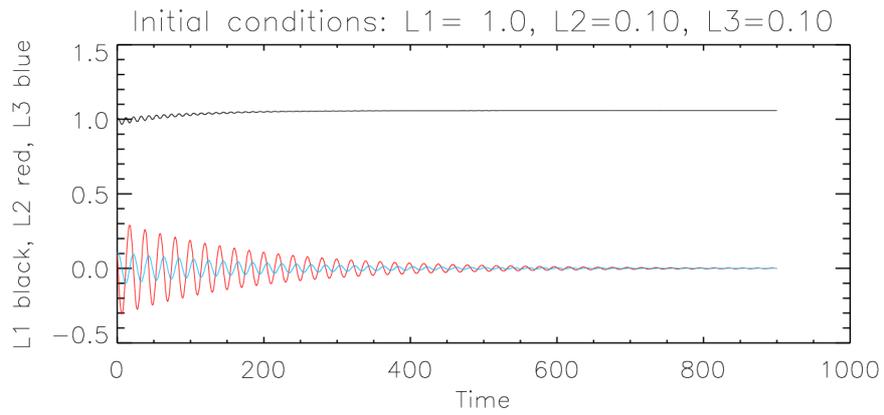

Figure 1. Time evolution of the three components of $\vec{L}$ in the MSM with $2kC'(L^2) = 0.1$ and with initial conditions $\vec{L}(0) = (1, 0.1, 0.1)^T$. The system relaxes to $\vec{L}_{(1)}$, see the text.

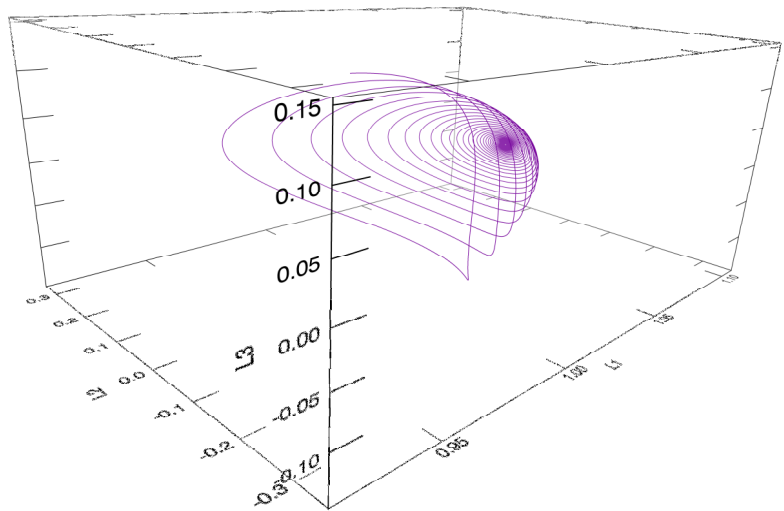

Figure 2. Phase portrait of the MSM, with initial conditions $\vec{L}(0) = (1, 0.1, 0.1)^T$, as in Figure 1.

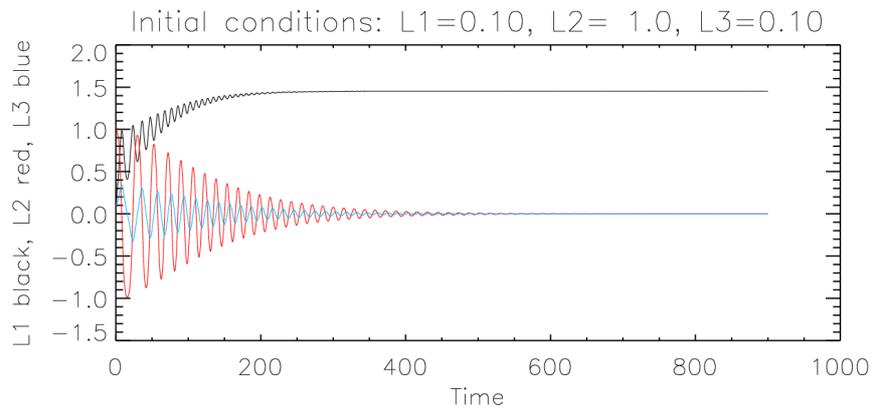

Figure 3. Time evolution of the three components of $\vec{L}$ in the MSM with $2kC'(L^2) = 0.1$ and with initial conditions $\vec{L}(0) = (0.1, 1, 0.1)^T$. The system relaxes to $\vec{L}_{(1)}$, see the text.

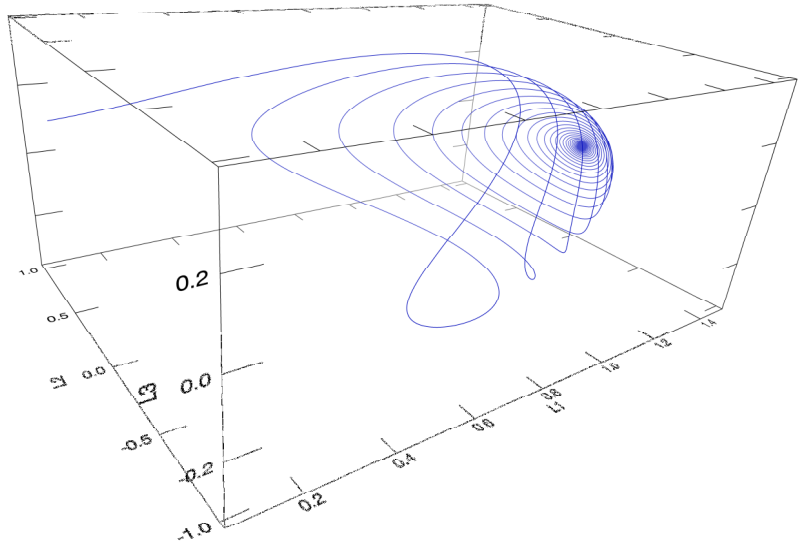

Figure 4. Phase portrait of the MSM, with initial conditions $\vec{L}(0) = (0.1, 1, 0.1)^T$, as in Figure 3.

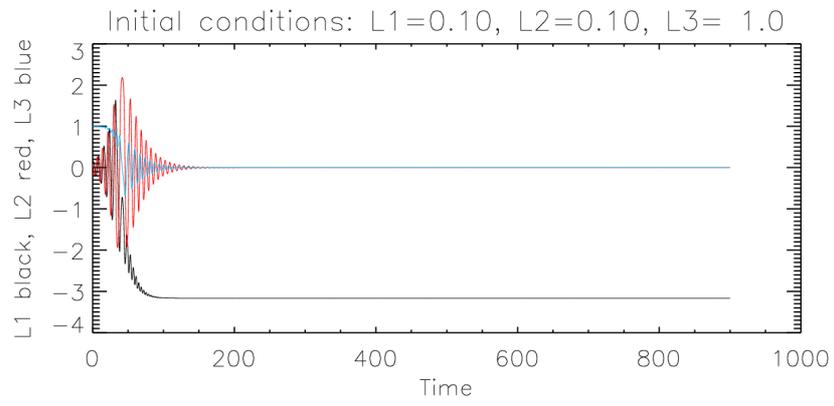

Figure 5. Time evolution of the three components of $\vec{L}$ in the MSM with $2kC'(L^2)=0.1$ and with initial conditions $\vec{L}(0)=(0.1,0.1,1)^T$. The system relaxes to $\vec{L}_{(1)}$, see the text.

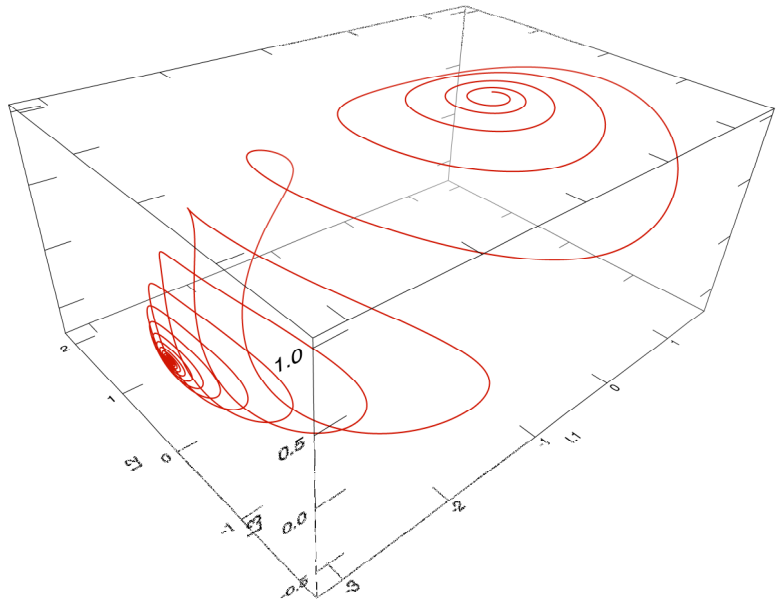

Figure 6. Phase portrait of the MSM, with initial conditions $\vec{L}(0)=(0.1,0.1,1)^T$, as in Figure 5.

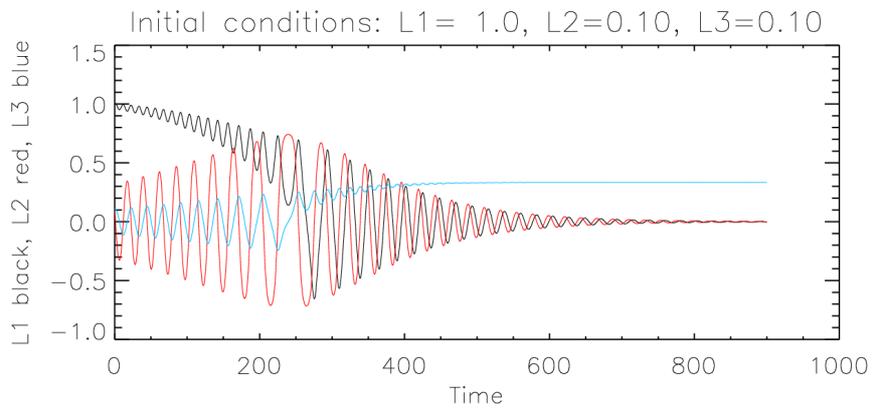

Figure 7. Time evolution of the three components of $\vec{L}$ in the MSM with $2kC'(L^2) = -0.1$ and with initial conditions $\vec{L}(0) = (1, 0.1, 0.1)^T$. The system relaxes to $\vec{L}_{(3)}$, see the text.

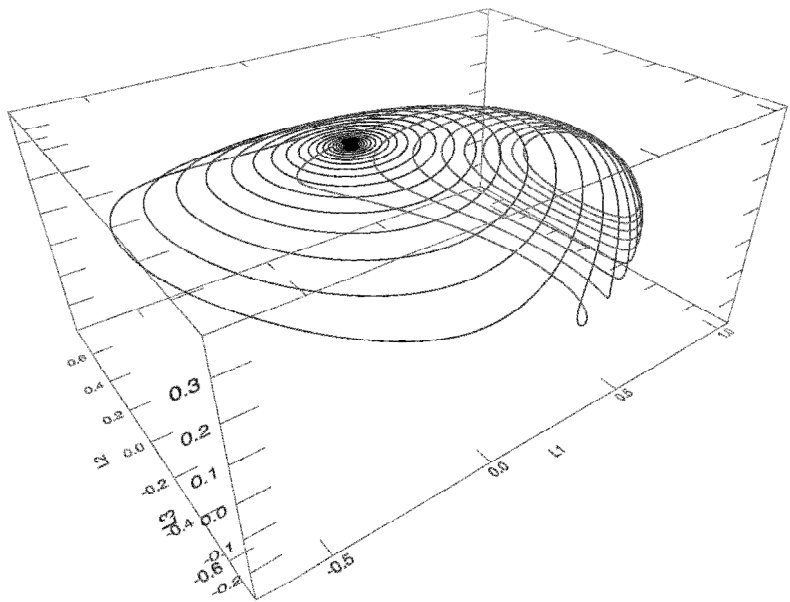

Figure 8. Phase portrait of the MSM, with initial conditions $\vec{L}(0) = (1, 0.1, 0.1)^T$, as in Figure 7.

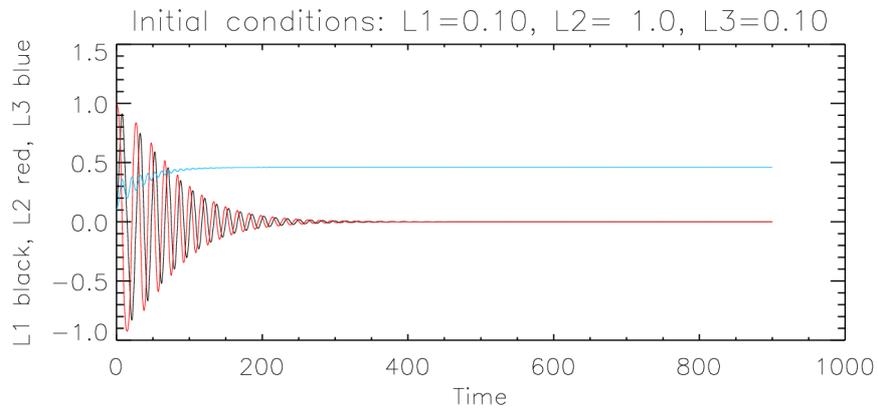

Figure 9. Time evolution of the three components of $\vec{L}$ in the MSM with $2kC'(L^2) = -0.1$ and with initial conditions $\vec{L}(0) = (0.1, 1, 0.1)^T$. The system relaxes to $\vec{L}_{(3)}$, see the text.

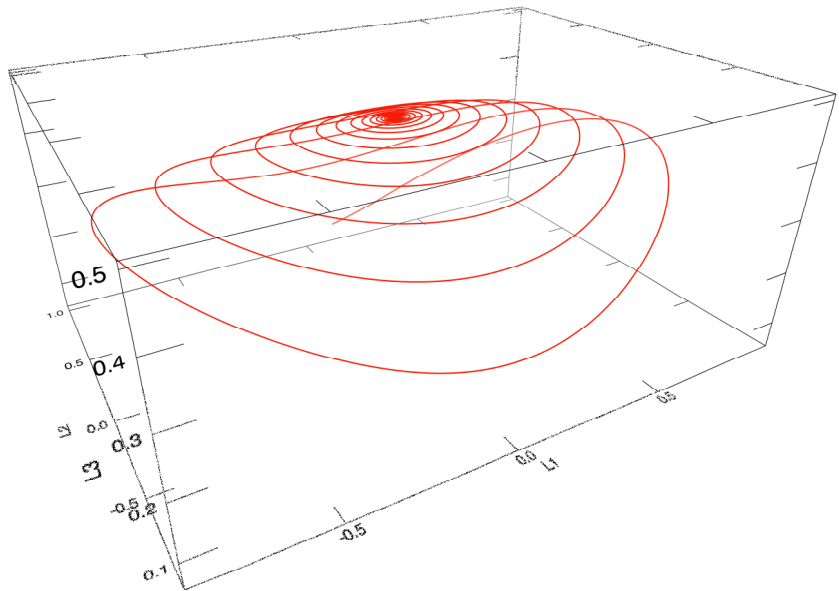

Figure 10. Phase portrait of the MSM, with initial conditions $\vec{L}(0) = (0.1, 1, 0.1)^T$, as in Figure 9.

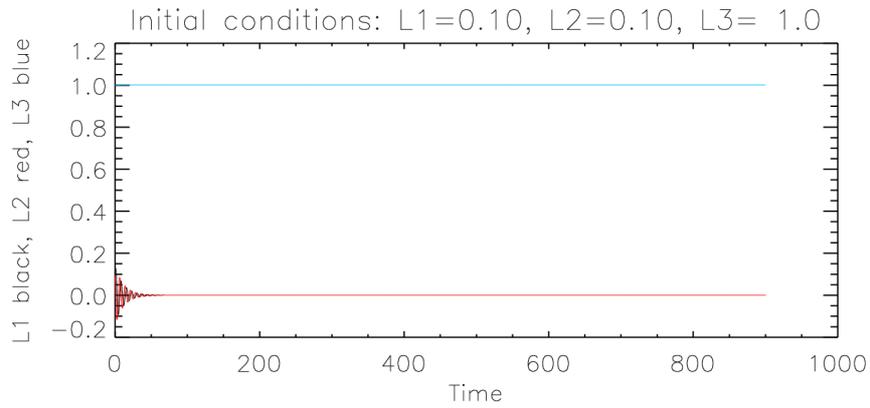

Figure 11. Time evolution of the three components of $\vec{L}$ in the MSM with $2kC'(L^2) = -0.1$ and with initial conditions $\vec{L}(0) = (0.1, 0.1, 1)^T$. The system relaxes to $\vec{L}_{(3)}$, see the text.

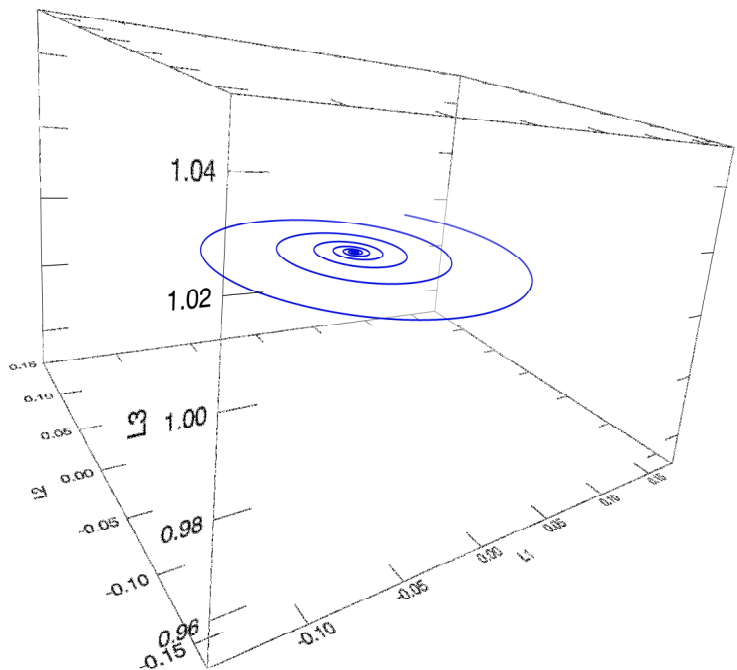

Figure 12. Phase portrait of the MSM, with initial conditions $\vec{L}(0) = (0.1, 0.1, 1)^T$, as in Figure 11.

Points of asymptotic equilibrium are 0-dimensional attractors, while most interesting dissipative systems (e.g. electric circuits, ecological systems, …) show *limit cycles* and higher dimensional (possibly strange) attractors. In the present example of the MSM a limit trajectory already exists in the $(\chi, \vec{p})$ phase space, simply the 1-dimensional attracting manifold corresponding to the equilibria (13) in the reduced phase space. The remaining part of this paper is dedicated to illustrating this point.

The Hamiltonian ODEs governing the motion of the variables $(\chi, \vec{p})$ read:

$$\begin{cases} \dot{\vec{\chi}} = A^T(\chi) \cdot \sigma \cdot A(\chi) \cdot \vec{p}, \\ \dot{\vec{p}} = -\vec{p}^T \cdot A^T(\chi) \cdot \sigma \cdot \dfrac{\partial A(\chi)}{\partial \chi} \cdot \vec{p} \end{cases} \quad (15)$$

(the triplet $\dot{\vec{\chi}}$ is an $SO(3)$-vector). In (15) the matrix $A(\chi)$ gives the linear, angle dependent and inertia dependent relationship between $\vec{p}$ and $\vec{\omega}$, as $\vec{\omega} = A(\chi) \cdot \vec{p}$. It can be expressed in terms of the relationship $\vec{\omega} = D(\chi) \cdot \dot{\vec{\chi}}$ between the angular velocity and the derivatives of Euler angles (see Figure 13):

$$D(\chi) = \begin{pmatrix} \cos\chi_3 & \sin\chi_1 \sin\chi_3 & 0 \\ -\sin\chi_3 & \sin\chi_1 \cos\chi_3 & 0 \\ 0 & \cos\chi_1 & 1 \end{pmatrix},$$

$$A(\chi) = \sigma^{-1} \cdot \left(D^{-1}(\chi)\right)^T.$$

Equations (15) *do correspond* to the equations of motion of $\vec{\omega}$ obtained by inserting $\vec{\omega}$ in (10): so, they have all the same solutions as those ones, re-expressed in the canonical variables.

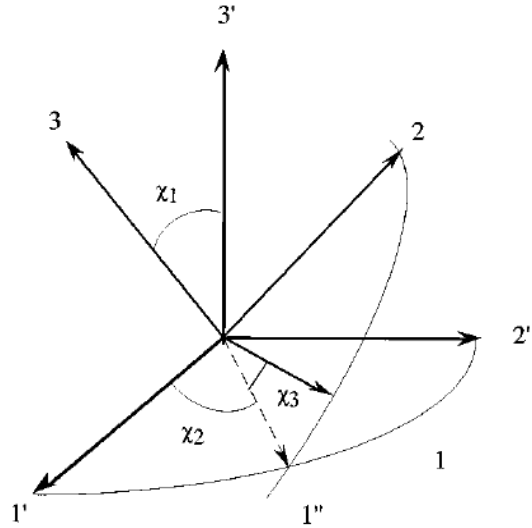

Figure 13. The Euler angles, used as Lagrangian variables for the rigid body.

As the MSM is added to this Hamiltonian system, in the space of the $\vec{\omega}$ vectors asymptotically stable points appear, that are translated into one-dimensional trajectories in the canonical variables $(\chi, \vec{p})$. These may be shown to be *limit cycles*. Let us consider in particular the angular velocity

$$\vec{\omega}^* = \Omega \hat{e}_2, \quad (16)$$

corresponding to the rigid body rotating around the $\hat{e}_2$ axis: if the momentum $I_2$ is either the largest or the smallest eigenvalue of $\sigma$, this $\vec{\omega}^*$ is an *asymptotically stable* point in the space of the angular velocities.

Placing (16) into (15), and considering $\vec{\omega} = A(\chi) \cdot \vec{p}$, we find that the corresponding velocities of the Euler angles read:

$$\vec{\omega} = \vec{\omega}^* \Rightarrow \begin{cases} \dot{\chi}_1 = -\Omega \sin \chi_3, \\ \dot{\chi}_2 = \Omega \dfrac{\cos \chi_3}{\sin \chi_1}, \\ \dot{\chi}_3 = -\Omega \dfrac{\cos \chi_1 \cos \chi_3}{\sin \chi_1}. \end{cases} \quad (17)$$

A possible, particularly readable, solution of (17), is

$$\begin{cases} \chi_1(t) = \dfrac{\pi}{2}, \\ \chi_2(t) = \Omega t + \chi_2(0), \\ \chi_3(t) = 0 \ \forall \ t. \end{cases} \quad (18)$$

The solution (18) represents *a uniform rotation around the node line*, that will coincide with the new *X* axis, being $\chi_3(t) = 0$ at every time. The value of $\vec{p}$ corresponding to $\vec{\omega}^*$ is the vector $\vec{p}^* = A^{-1}(\chi^*) \cdot \vec{\omega}^*$, with $\chi^*(t) = (\frac{\pi}{2}, \Omega t + \chi_2(0), 0)$, that reads

$$\vec{p}^* = \begin{pmatrix} 0 \\ I_2 \Omega \\ 0 \end{pmatrix}. \quad (19)$$

This $\vec{p}^* = A^{-1}(\chi^*) \cdot \vec{\omega}^*$ and the aforementioned $\chi^*(t)$ form a solution of the ODEs (15) in the space of canonical variables $(\chi, \vec{p})$, as they correspond to the $\vec{\omega}^*$ solving the Hamilton equations in the $\vec{\omega}$-space. What happens to this trajectory when the MSM is at work? When this is so, a new term appears in the ODE for $\vec{p}$, i.e. some $\vec{\Delta}_p(\chi, \vec{p})$ representing *the same effect* of the metriplectic term $\vec{\Delta}_\omega(\vec{\omega})$ on the rigid body motion, now written in the canonical variables $(\chi, \vec{p})$. In order to find the analytical expression of $\vec{\Delta}_p(\chi, \vec{p})$, one simply imposes that the ODEs for $\vec{\omega}$ are *the same* in the two systems of variables. Under the assumption that the ODEs of the angles *are not affected by any metric contribution*, i.e. $\vec{\Delta}_\chi(\chi, \vec{p}) = 0$, this reasoning leads to:

$$\vec{\Delta}_p = 2\zeta C' A^{-1} \cdot \Gamma \cdot \sigma^2 \cdot A \cdot \vec{p}, \quad (20)$$

where all the dependencies on $(\chi, \vec{p})$ are omitted. The foregoing term must be added to the Hamiltonian contribution in (15), becoming:

$$\begin{cases} \dot{\vec{\chi}} = D^{-1} \cdot A \cdot \vec{p}, \\ \dot{\vec{p}} = -\vec{p}^{\mathrm{T}} \cdot A^{\mathrm{T}} \cdot \sigma \cdot \dfrac{\partial A}{\partial \chi} \cdot \vec{p} + \\ \qquad + 2\zeta C' A^{-1} \cdot \Gamma \cdot \sigma^2 \cdot A \cdot \vec{p}. \end{cases} \qquad (21)$$

It is possible to see that the curve (18) and (19) in the full canonical phase space of the rigid body solves the total (21), because the metriplectic term $\vec{\Delta}_p = 2\zeta C' A^{-1} \cdot \Gamma \cdot \sigma^2 \cdot A \cdot \vec{p}$ can be shown to vanish in the correspondence of $(\chi^*(t), \vec{p}^*)$. This follows because the matrix

$$\Gamma(\vec{\omega}^*) = \frac{k\Omega^2}{\zeta} \begin{pmatrix} 1 & 0 & 0 \\ 0 & 0 & 0 \\ 0 & 0 & 1 \end{pmatrix}$$

has the vector

$$\sigma^2 \cdot A(\chi^*) \cdot \vec{p}^* = \begin{pmatrix} 0 \\ I_2^2 \Omega \\ 0 \end{pmatrix}$$

in its kernel. Hence, one may state that, *once on this orbit, the rigid body will remain there even in the presence of the MSM*, the torque of which indeed vanishes along this trajectory.

As mentioned before, provided that $I_2$ is either the maximum or the minimum of the eigenvalues of $\sigma$, the point (16) represents a stable equilibrium point for the Hamiltonian free rigid body, and an asymptotically stable point for its counterpart with the MSM. If the canonical variables $(\chi, \vec{p})$ are adopted, the corresponding $(\chi^*(t), \vec{p}^*)$ is *an attracting limit cycle* of the metriplectic system. In order to check this, one expands in $(\delta\chi, \delta\vec{p})$ around $(\chi^*(t), \vec{p}^*)$ the equations (21). After this expansion, the perturbations $(\delta\chi_2, \delta p_2)$ decouple from the other ones

$$\delta\dot{\chi}_2 = \frac{\delta p_2}{I_2}, \quad \delta\dot{p}_2 = 0,$$

resulting in a perturbation that will affect the cycle as:

$$\chi_2(0) \mapsto \chi_2(0) + \delta\chi_2(0), \quad \Omega \mapsto \Omega + \frac{\delta p_2(0)}{I_2}$$

(*the same cycle*, but swept by *starting from another point* and with *another velocity*). The other variables evolve according to a system of linear ODEs coupled among themselves:

$$\begin{cases} \delta\dot{\chi}_1 = -\frac{(I_1-I_2)\Omega}{I_1}\delta\chi_3 + \frac{1}{I_1}\delta p_1, \\ \delta\dot{\chi}_3 = \Omega\delta\chi_1 + \frac{1}{I_3}\delta p_3, \\ \delta\dot{p}_1 = I_2\Omega^2\delta\chi_1 + 2\beta I_2(I_1^2 - I_2^2)\Omega^3\delta\chi_3 + \\ \quad + 2\beta(I_1^2 - I_2^2)\Omega^2\delta p_1 + \Omega\delta p_3, \\ \delta\dot{p}_3 = \frac{I_2(I_1-I_2)\Omega^2}{I_1}\delta\chi_3 + \frac{(I_1-I_2)\Omega}{I_1}\delta p_1 + \\ \quad + 2\beta(I_3^2 - I_2^2)\Omega^2\delta p_3, \end{cases} \quad (22)$$

where the assumption $\beta = kC'(L^2(\vec{\chi}^*, \vec{p}^*))$ has been made. The study of the system (22) gives the expected results, viz. the perturbations of the limit cycle (18) and (19) tend to zero with time as $I_1 - I_2$ and $I_1 - I_3$ have the same sign, i.e. as the body nearly rotates around a principal axis with either maximum or minimum moment of inertia. This properly represents how the point-like equilibrium (13) in the reduced phase space turns into an attractive limit cycle in the complete canonical phase space.

**4 Conclusions**

In this paper we have briefly reviewed the concept of complete metriplectic systems, the extension of Hamiltonian systems possessing asymptotically stable equilibria. Such systems may represent energy-conserving, entropy increasing dynamics, particularly useful for describing relaxation processes.

Turning non-Hamiltonian systems into dynamics described by Leibniz bracket algebrae can be usefully applied to systems showing finite-dimensional attractors, as limit cycles. This is indeed the case of interesting systems in applied physics, space physics and geophysics, biophysics or mathematical ecology.

Here an example of a complete metriplectic system showing limit cycle attracting orbits is reported: a free rigid body to which a suitable external torque is applied, able to modify the angular momentum, and angular velocity, without changing the energy. Such a mechanism, referred to as metriplectic servo-motor, may drag the angular velocity of a free rotator to align along one of its principal axes of inertia, without dissipation of mechanical energy.

Such a final configuration, obtained via a MSM, is an asymptotically stable, point-like equilibrium in the space of angular velocity (or angular momentum), but corresponds to a 1-dimensional limit cycle when the system is described via Euler angles and their respective canonically conjugate momenta.

**Acknowledgements**
PJM received support from the US Dept.of Energy Contract DE-FG05-80ET-53088 and from a Forschungspreis from the Alexander von Humboldt Foundation. He would like to warmly acknowledge the hospitality of the Numerical Plasma Physics Division of the Max Planck IPP, Garching, Germany.